# Strong localization of oxidized $Co^{3+}$ state in cobalt-hexacyanoferrate

Hideharu Niwa[1,2*], Masamitsu Takachi[2], Jun Okamoto[3], Wen-Bin Wu[3], Yen-Yi Chu[3], Amol Singh[3], Di-Jing Huang[3], Yutaka Moritomo[1,2,4*]

[1]Faculty of Pure and Applied Science, University of Tsukuba, Tsukuba 305-8571, Japan

[2]Graduate School of Pure and Applied Science, University of Tsukuba, Tsukuba 305-8571, Japan

[3]National Synchrotron Radiation Research Center, Hsinchu 30076, Taiwan

[4]Tsukuba Research Center for Energy Materials Science (TREMS), University of Tsukuba, Tsukuba 305-8571, Japan

Contact information:

Yutaka Moritomo / Hideharu Niwa

Faculty of Pure and Applied Science,

Univ. of Tsukuba, Tennodai 1-1-1, Tsukuba 305-8571, Japan

Tel +81-29-853-4337 / +81-29-853-4216

e-mail: moritomo.yutaka.gf@u.tsukuba.ac.jp / niwa.hideharu.ga@u.tsukuba.ac.jp

Secondary batteries are important energy storage devices for a mobile equipment, an electric car, and a large-scale energy storage. Nevertheless, variation of the local electronic state of the battery materials in the charge (or oxidization) process are still unclear. Here, we investigated the local electronic state of cobalt-hexacyanoferrate ($Na_xCo[Fe(CN)_6]_{0.9}$), by means of resonant inelastic X-ray scattering (RIXS) with high energy resolution (~100 meV). The $L$-edge RIXS is one of the most powerful spectroscopic technique with element- and valence-selectivity. We found that the local electronic state around $Co^{2+}$ in the partially-charged $Na_{1.1}Co^{2+}_{0.5}Co^{3+}_{0.5}[Fe^{2+}(CN)_6]_{0.9}$ film ($x = 1.1$) is the same as that of the discharged $Na_{1.6}Co^{2+}[Fe^{2+}(CN)_6]_{0.9}$ film ($x = 1.6$) within the energy resolution, indicating that the local electronic state around $Co^{2+}$ is invariant against the partial oxidization. In addition, the local electronic state around the oxidized $Co^{3+}$ is essentially the same as that of the fully-charged film $Co^{3+}[Fe^{2+}(CN)_6]_{0.3}[Fe^{3+}(CN)_6]_{0.6}$ ($x = 0.0$) film. Such a strong localization of the oxidized $Co^{3+}$ state is advantageous for the reversibility of the redox process, since the localization reduces extra reaction within the materials and resultant deterioration.

Lithium-ion/sodium-ion secondary batteries (LIBs/SIBs) are important energy storage devices for a mobile equipment, an electric car, and a large-scale energy storage. The device stores electric energy as material energy through a reversible redox process in cathode and anode materials. To comprehend what happens in battery materials in the charge (oxidization) process, we should know variation of the local electronic state in a valence-selective manner. In other words, we should clarify how far the effect of the oxidized site spreads and what kind of electronic state the oxidized site is.

Among the cathode materials, the metal ($M$) - hexacyanoferrates ($Na_xM[Fe(CN)_6]_y$ [1]) are attracting current interest of material scientists, because they are promising cathode materials for LIBs[2-4] and SIBs.[5-16] The $M$-hexacyanoferrates consist of three-dimensional jungle-gym type -$M$-NC-Fe-CN-$M$-NC- Fe-CN-$M$-NC- network and $Na^+$ and $H_2O$, which are accommodated in the network nanopores. Most of the $M$-HCFs show the face-centered cubic ($Fm\bar{3}m$; $Z = 4$). Figure 1 shows schematic illustration of the redox process in cobalt-hexacyanoferrate. The constituent Co ions take the divalent high-spin (HS) state in the discharge state. In the charge (oxidization) process, the Co sites are selectively oxidized with deintercalation of $Na^+$. Eventually, all the Co sites are oxidized in the charge state, where the Co sites take trivalent low-spin (LS) state.[5] The charge state is unstable, and hence, the discharge (reduction) process spontaneously takes place when the battery are connected to an external load.

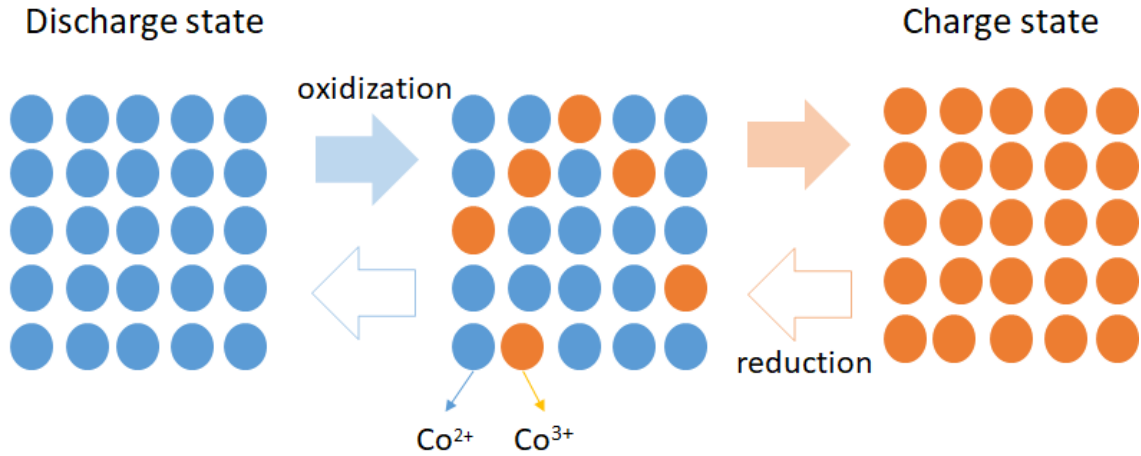

Figure 1: Schematic illustration of redox process in cobalt-hexacyanoferrate.

In the actual compounds, the charge process accompanies significant structural change, *e.g.*, volume expansion/shrinkage. For example, the lattice constant ($a$) of $Na_xCo[Fe(CN)_6]_{0.9}$ steeply decreases with charge process from 10.2 Å at $x$ = 1.6 to 9.9 Å at $x$ =0.8, because the ionic radius ($r_{HS}$= 0.745 Å) of HS $Co^{2+}$ is much larger than that ($r_{LS}$ = 0.545 Å) of LS $Co^{3+}$. Such a severe structural change is considered to influence the local electronic states. We note that $Na_xCo[Fe(CN)_6]_{0.9}$ has considerable $Fe(CN)_6$ vacancies, where $H_2O$ molecules coordinate to Co instead of CN. Takachi *et al.*[5] systematically investigated the structural and electronic properties of $Na_xCo[Fe(CN)_6]_{0.9}$ against $Na^+$ concentration ($x$). The crystal structure remains face-centered cubic in the whole region of $x$ (0.0 < $x$ < 1.6) without showing phase separation nor phase transition. This suggests that the oxidized $Co^{3+}$ sites are uniformly distributed. The X-ray absorption spectroscopy (XAS) around the Co K-edge suggests coexistence of HS $Co^{2+}$ and LS $Co^{3+}$ in the region of 0.6 < $x$ < 1.6. Further analyses of the XAS, however, are impossible due to lack of valence-selectivity. The *L*-edge resonant inelastic X-ray scattering (RIXS) with high energy resolution (~100 meV) enable us to detect even a slight variation of the local electronic state around the Co site in a valence-selective manner.

Here, we investigated how the local electronic state around the Co cite changes in the charge process in Na$_x$Co[Fe(CN)$_6$]$_{0.9}$ by means of the Co $L_3$-edge RIXS with high energy resolution. The RIXS revealed that the local electronic state around Co$^{2+}$ is invariant within the energy resolution against the partial oxidization. In addition, the local electronic state around the oxidized Co$^{3+}$ is essentially the same as that of the fully-charged film ($x = 0.0$). The localization of the oxidized Co$^{3+}$ state, which is probably stabilized by the heterogeneous lattice structure, is advantageous for the reversibility of the redox process.

**X-ray absorption spectra around the Co L$_3$-edge**

We prepared three Na$_x$Co[Fe(CN)$_6$]$_{0.9}$ films with different Na$^+$ concentration ($x$) by means of the electrochemical method. In Table 1, we listed the $x$-controlled Na$_x$Co[Fe(CN)$_6$]$_{0.9}$ films together with the nominal valence state of Co and Fe. For convenience of explanation, we will call the films as the HS Co$^{2+}$ ($x = 1.6$), mixed ($x = 1.1$), and LS Co$^{3+}$ ($x = 0.0$) films, respectively.

| name | $x$ | nominal valence state | chare/discharge state |
|---|---|---|---|
| HS Co$^{2+}$ | 1.6 | Na$_{1.6}$Co$^{2+}$[Fe$^{2+}$(CN)$_6$]$_{0.9}$ | Discharged |
| mixed | 1.1 | Na$_{1.1}$Co$^{2+}_{0.5}$Co$^{3+}_{0.5}$[Fe$^{2+}$(CN)$_6$]$_{0.9}$ | partially-charged |
| LS Co$^{3+}$ | 0.0 | Co$^{3+}$[Fe$^{2+}$(CN)$_6$]$_{0.3}$[Fe$^{3+}$(CN)$_6$]$_{0.6}$ | Charged |

Table 1: List of the $x$-controlled Na$_x$Co[Fe(CN)$_6$]$_{0.9}$ films with nominal valence state.

Figure 2 shows absorption spectra around the Co $L_3$-edge of the three films. The measurements were performed at Taiwan Light Source (TLS) BL08B1 beamline at the National Synchrotron Radiation Research Center (NSRRC) in Taiwan. The spectra indicated by solid and dashed curves were obtained in the total electron yield (TEY) and partial fluorescence yield (PFY) modes, respectively. The TEY mode is surface-sensitive. In all the films, the peak features of the TEY spectra are similar to the PFY spectra.

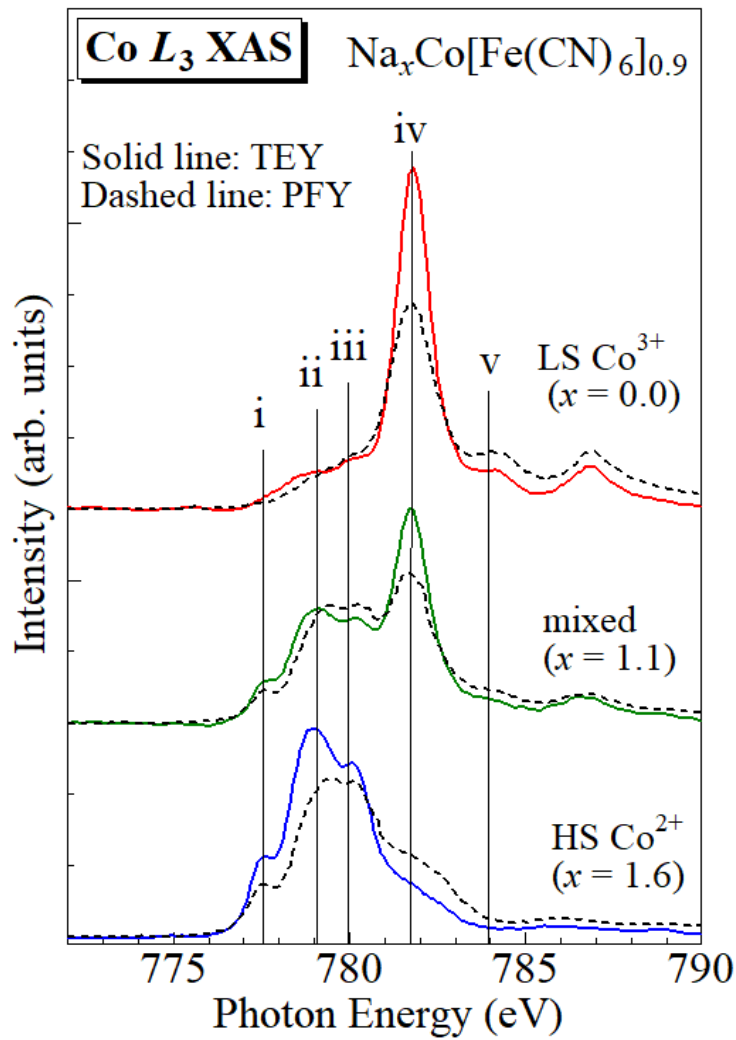

Figure 2: X-ray absorption spectra around the Co $L_3$-edge of the Na$_x$Co[Fe(CN)$_6$]$_{0.9}$ films against $x$. Solid and dashed lines represent the TEY and PFY spectra, respectively. Vertical lines (i – v) represent the incident photon energies in the RIXS measurements (*vide infra*).

In the HS $Co^{2+}$ film ($x$ = 1.6), the TEY spectrum shows characteristic peaks around 777.5, 779, and 780 eV. The spectral feature is similar to that of CoO which serves as HS $Co^{2+}$ reference.[17] In the LS $Co^{3+}$ film ($x$ = 0.0), the TEY spectrum consists of sharp peak at higher energy around 782 eV with a shoulder structure around 784 eV. The spectral feature is similar to that of $EuCoO_3$ which serves as LS $Co^{3+}$ reference.[17] The TEY spectrum of the mixed film ($x$ = 1.1) is close to the superimposed spectrum of those of the HS $Co^{2+}$ and LS $Co^{3+}$ films. This suggests coexistence of the HS $Co^{2+}$ and LS $Co^{3+}$ sites in the mixed film. In Fig. 2S, we show the absorption spectra around the Co $L_{2,3}$-edge of the three films.

**RIXS spectra around the Co $L_3$-edge**

Figure 3 shows the RIXS spectra around the Co $L_3$-edge of the three films: (a) HS $Co^{2+}$ ($x$ = 1.6), (b) mixed ($x$ = 1.1), and (c) LS $Co^{3+}$ ($x$ = 0.0). The measurements were performed at TLS BL05A1 beamline [18] at the NSRRC in Taiwan. The horizontal axis represents the energy loss of the scattered X-ray, which are mainly transferred to the crystal-field excitations of Co. The excitation photon energies ($E_{ex}$) are indicated by vertical lines (i – v) in Fig. 2. The spectra were normalized to the incident photon flux. As discussed above, the RIXS spectra at $E_{ex}$ < 779.8 eV are dominated by the scattering due to HS $Co^{2+}$ while the spectra at $E_{ex}$ > 781.7 eV are dominated by the scattering due to LS $Co^{3+}$. Actually, at $E_{ex}$ = 777.4 and 779.0 eV, the RIXS spectra of the HS $Co^{2+}$ film [(a)] is much stronger than those of the LS $Co^{3+}$ film [(c)]. We show magnified spectra in Fig. S4.

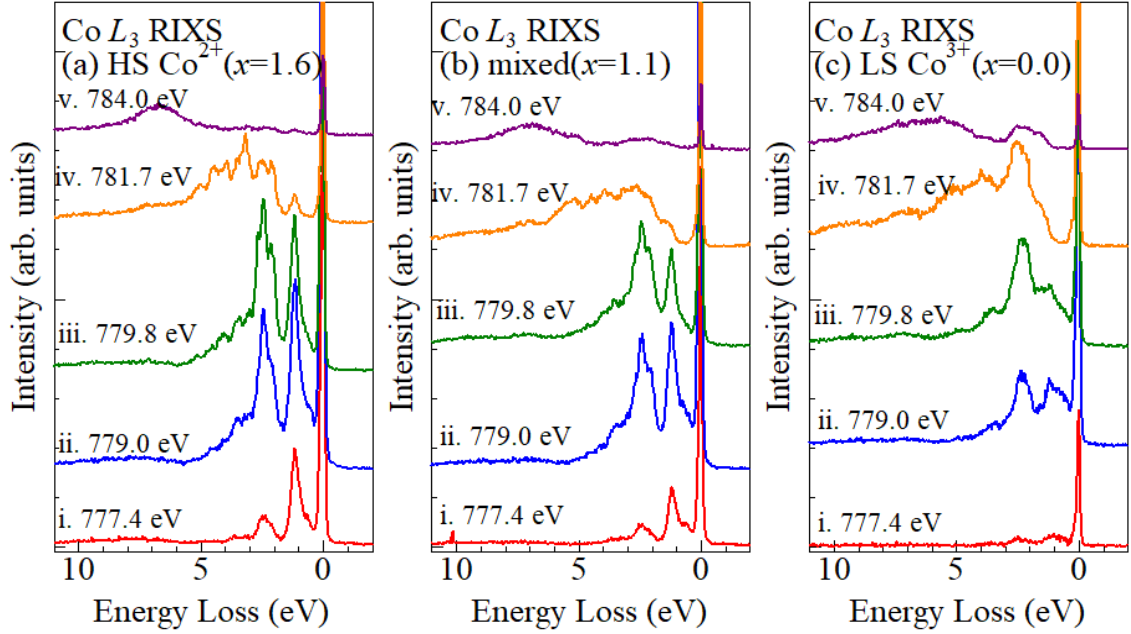

Figure 3: RIXS spectra around the Co $L_3$-edge of the Na$_x$Co[Fe(CN)$_6$] films: (a) $x$ = 1.6, (b) 1.1, and (c) 0.0. The spectra were normalized to the incident photon flux.

In the HS Co$^{2+}$ film [(a)], the spectra at $E_{ex}$ = 777.4 , 779.0 and 779.8 eV show two intense peaks around 1.1 and 2 -3 eV. The spectral feature is similar to the Co $L_3$-edge RIXS spectra of HS Co$^{2+}$ in CoO.[19,20] In CoO, the peak around 1.1 eV is due to excitations to the $^4T_{2g}(^4F)$ states, the shoulder around 2.2 eV is due to transitions to the $^4A_{2g}(^4F)$ states, and the manifold of peaks around 2.5 – 3.0 eV are mainly due to transitions to the $^4T_{1g}(^4P)$ states.[19] The crystal field value (10$Dq$), which is the energy difference between the $^4T_{2g}(^4F)$ and $^4A_{2g}(^4F)$ states, is evaluated to be 0.95 eV in the HS Co$^{2+}$ film. At 784.0 eV, broad band at ~7 eV is probably due to the charge-transfer excitations from CN$^-$ to Co$^{2+}$. In the LS Co$^{3+}$ film [(c)], the spectra at $E_{ex}$ =781.7 eV shows intense peaks around ~ 2.3 eV with a shoulder structure around 3.0 – 6.0 eV. The spectral feature is similar to the Co $L_3$-edge RIXS spectra of LS Co$^{3+}$ in LaCoO$_3$,[21,22] but its crystal-field excitation energy is much larger.[22] Tomiyasu *et al.* reported Co $L_3$-edge RIXS of LaCoO$_3$ single crystal with high energy resolution (~ 80 meV). In LaCoO$_3$, the RIXS spectrum at 20 K shows intense peak around 1.3 eV with a shoulder structure around 0.6 eV. In the LS $Co^{3+}$ film[(c)], at $E_{ex}$ = 718.7 eV, the corresponding features are observed around 2.3 and 1.8 eV. Hereafter, we will refer the RIXS spectra of the HS $Co^{2+}$ [(a)] and LS $Co^{3+}$ [(c)] films as $\phi_{2+}$ and $\phi_{3+}$, respectively. In the mixed film [(b)], at $E_{ex}$ < 779.8 eV, the spectra shows two-peak feature, which is characteristic to $\phi_{2+}$ [(a)]. At 784 eV, the spectrum shows a broad band around 1.0 – 3.0 eV and ~ 7 eV, whose feature is close to $\phi_{3+}$ [(c)].

In order to quantitatively analyze the RIXS spectra ($\phi_{mixed}$) of the mixed film ($x$ = 1.1), we compared $\phi_{mixed}$ with linear combination of $\phi_{2+}$ and $\phi_{3+}$. We performed least-squares fitting of $\phi_{mixed}$ with a trial function: $c\phi_{2+} + (1 - c) \phi_{3+}$. The adjustable parameter ($c$) were 0.48 at (a) 777.4 eV, 0.62 at (b) 779.0 eV, 0.52 at (c) 779.8 eV, 0.44 at (d) 781.7 eV, and 0.77 at (e) 784.0 eV. Except at (e) 784.0 eV, the $c$ values are close to 1/2. Figure 4 shows comparison of $\phi_{mixed}$ with ($\phi_{2+}+\phi_{3+}$)/2 at the respective $E_{ex}$. The ($\phi_{2+}+\phi_{3+}$)/2 spectra quantitatively reproduces $\phi_{mixed}$. Especially, the agreement is good at (a) 777.4 eV, (b) 779.0 eV, and (c) 779.8 eV, where the spectra are dominated by scattering due to HS $Co^{2+}$. Utilizing the characteristics of high energy resolution, we evaluated the energy shifts ($\Delta E$) of the peaks around 1.1 and 2 -3 eV between $\phi_{2+}$ and $\phi_{mixed}$: $\Delta E$ = 0.07 and 0.00 eV at (a) 777.4 eV, 0.04 and 0.05 eV at (b) 779.0 eV, and 0.00 and 0.02 eV at (c) 779.8 eV. Thus, the local electronic state around HS $Co^{2+}$ is invariant within the energy resolution (< 100 meV) against the partial oxidization. At (d) 781.7 eV and (e) 784.0 eV, where the spectra at are dominated by scattering due to LS $Co^{3+}$, the agreement is satisfactory except for slight difference in peak energy and intensity. At (e) 784.0 eV, the broad band around ~ 2.3 eV is well reproduce by $\phi_{3+}$. This indicates that the oxidized $Co^{3+}$ site takes LS configuration because the ~2.3 eV peak is characteristic to the crystal-filed excitation of the LS $Co^{3+}$.[22] The slight spectral difference is perhaps due to the partial oxidization of $Fe^{2+}$ in the LS $Co^{3+}$ ($x$ = 0.0) film. In Fig. 3S, we show X-ray absorption spectra around the Fe $L_{2,3}$-edge. The $x$ =0.0 spectrum is significantly different from the $x$ = 1.6 and 1.1 spectra, reflecting the partial oxidization of $Fe^{2+}$.

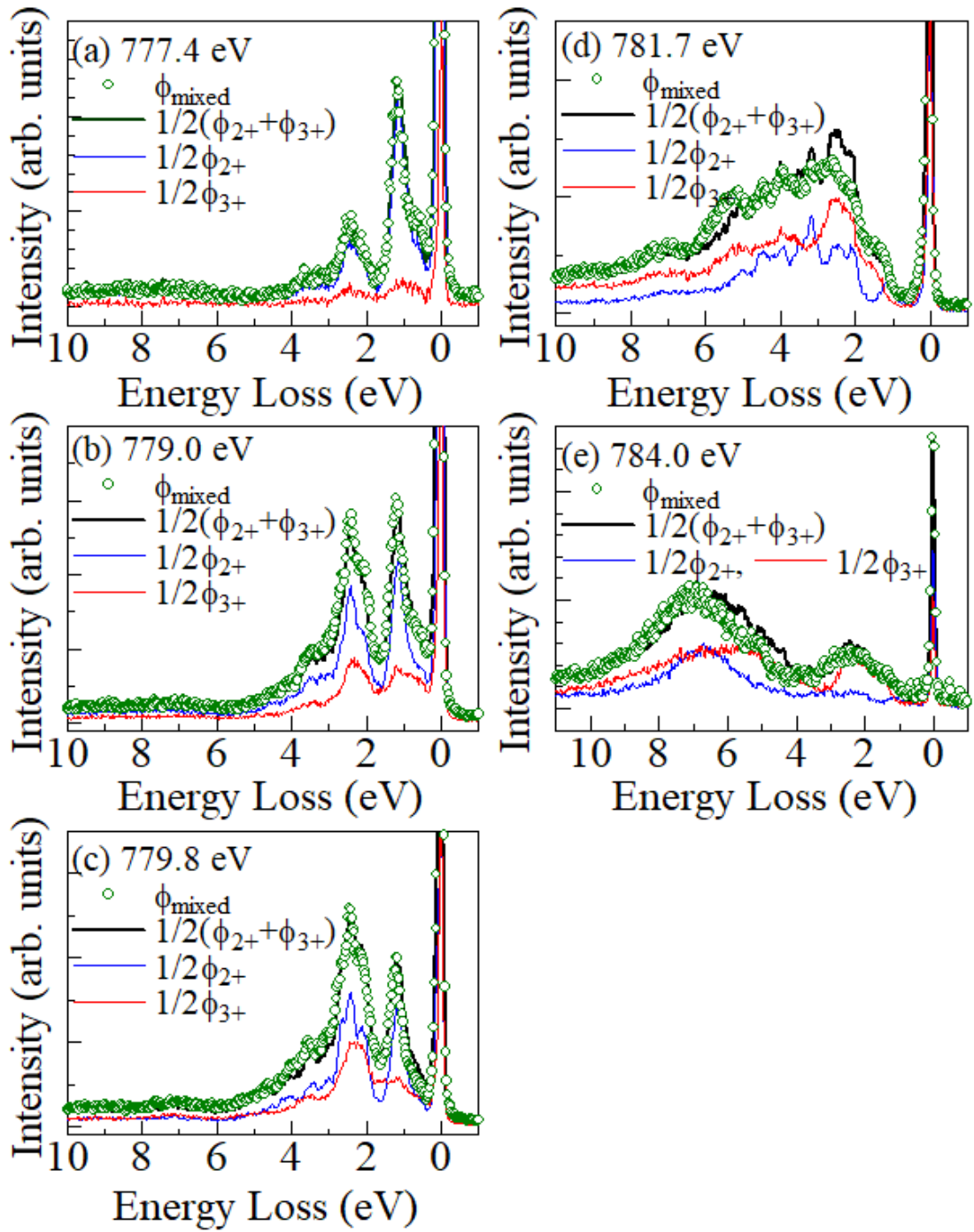

Figure 4: RIXS spectra ($\phi_{mixed}$) at $x = 1.1$ around the Co $L_3$-edge at (a) 774.0 eV, (b) 779.0 eV, (c) 779.8 eV, (d) 781.7 eV, and (e) 784.0 eV. $\phi_{2+}$ and $\phi_{3+}$ are the corresponding RIXS spectra of the $x = 1.6$ and 0.0 films, respectively. $\phi_{mixed}$, $\phi_{2+}$, and $\phi_{3+}$ were normalized to the incident photon flux. The thick black curves represent $(\phi_{2+}+\phi_{3+})/2$.

**Discussion**

Now, let us discuss the local electronic state around the Co site in the mixed ($x = 1.1$) film. The RIXS spectroscopy revealed (1) the local electronic state around HS $Co^{2+}$ is invariant within the energy resolution (~100 meV) and (2) the oxidized $Co^{3+}$ site takes the LS configuration in the charge process. In the discharge state, all the Co sites take the HS $Co^{2+}$ configuration. In the oxidization process, an electron is removed from a Co site to produce an oxidized $Co^{3+}$ state among the inherent HS $Co^{2+}$ environment. The RIXS spectroscopy clearly indicate that the effect of the oxidized site does not reach to the neighboring $Co^{2+}$ site, because the local electronic state around the $Co^{3+}$ site is invariant against partial oxidization. On the other hand, the oxidized $Co^{3+}$ site takes the LS configuration, since the ligand field (crystal-field) is much stronger around the $Co^{3+}$ site. In other words, the size of the $Co^{3+}N_6$ octahedron should be much smaller than that of $Co^{2+}N_6$. These arguments reaches a picture of the partially-oxidized state, *i.e.*, the electronic state is strongly localized within the range of each Co atom by making the lattice structure heterogeneous at the atomic level.

Why such a heterogeneous structure is possible in $Na_xCo[Fe(CN)_6]_{0.9}$? We ascribed the heterogeneity to the structural flexibility in the 3D network, -Co-NC-Fe-CN-Co-. The network is fairly sparse and has 10% $Fe(CN)_6$ vacancies. Actually, the density (~ 1.9 g/cm$_3$) of $Na_xCo[Fe(CN)_6]_{0.9}$ is much smaller as compared with that (= 5.0 g/cm$^3$) of layered oxides. With such a structure, the local compression around the oxidized $Co^{3+}$ site in the -Co-NC-Fe-CN-Co-V-Co-NC- (V represents the $[Fe(CN)_6]$ vacancy) chain is well compensated by expansion of the space at the vacancy site. Thus, the strong localization of the oxidized $Co^{3+}$ state is stabilized by the $[Fe(CN)_6]$ vacancies. The localization of the oxidized state is advantageous for the reversibility of the redox process, since the localization reduces extra reaction within the materials and resultant deterioration.

**Summary**

The $L$-edge RIXS investigation of the $Na_xCo[Fe(CN)_6]_{0.9}$ films revealed (1) the local electronic state around HS $Co^{2+}$ is invariant within the energy resolution (~100 meV) and (2) the oxidized $Co^{3+}$ site takes LS configuration in the charge process. We ascribed the strong localization of the oxidized $Co^{3+}$ state to the heterogeneous lattice structure. The localization of the oxidized state is advantageous for the reversibility of the redox process, since the localization reduces extra reaction within the materials and resultant deterioration. Thus, the advanced spectroscopy with use of 3$^{rd}$ generation synchrotron-radiation facility is a powerful tool to reveal the microscopic process within the battery materials.

**Method**

**Preparation and characterization of Na$_{1.6}$Co[Fe(CN)$_6$]$_{0.9}$ film**

The electrochemical deposition of the Na$_{1.6}$Co[Fe(CN)$_6$]$_{0.9}$ film was performed in a three-pole beaker-type cell. The working, counter, and standard electrodes were an indium tin oxide (ITO) transparent, Pt, and standard Ag/AgCl electrodes, respectively. The electrolyte was an aqueous solution containing 0.8 mmol/L K$_3$[Fe(CN)$_6$], 0.5 mmol/L Co(NO$_3$)$_2$, and 5.0 mol/L NaNO$_3$. The films were deposited on the ITO electrode under potentiostatic conditions at -0:45V vs. the Ag/AgCl electrode. The thickness of the film was around 1.4 μm, which was determined by a profilometer (Deltak-3030). The chemical composition of the film was determined by the inductively coupled plasma (ICP) method and CHN organic elementary analysis (PerkinElmer 2400 CHN Elemental Analyzer). The compound contains crystal waters as Na$_{1.6}$Co[Fe(CN)$_6$]$_{0.9}$2.9 H$_2$O. The X-ray diffraction patterns of the Na$_{1.6}$Co[Fe(CN)$_6$]$_{0.9}$ films were obtained with a Cu K$\alpha$ lines. All the reflections can be indexed with the face-centered cubic structure. The lattice constants (*a*) were 10.27 Å. The scanning electron microscopy (SEM) revealed that the films consist of crystalline grains of several hundred nm in diameter.[23]

**Battery cell**

The *x* value of the Na$_x$Co[Fe(CN)$_6$]$_{0.9}$ film was finely controlled in a beaker-type cell with two-pole configuration. The cathode, anode, and electrolyte were the film , Na metal, and M NaClO$_4$ in propylene carbonate (PC), respectively. The electrochemical control was performed with a potentiostat (HokutoDENKO HJ1001SD8) in an Ar-filled glove box. The active areas of the films were about 0.5 cm$^2$. The cut-off voltage was in the range of 2.0 to 4.0 V. The charge rate was about 1 C. Figure S1 shows charge curve of the Na$_x$Co[Fe(CN)$_6$]$_{0.9}$ film. The mass of each film was evaluated from thickness, area, and ideal density. The *x* value was evaluated from the total current under the assumption that *x* = 1.6 (0.0) is in the discharge (charge) state. Thus prepared films are listed in Table 1.

**X-ray absorption spectra around the Co $L_{2,3}$- and Fe $L_{2,3}$-edges**

The XAS measurements were conducted at TLS BL08B1 beamline at the NSRRC in Taiwan. The absorption spectra around the Co $L_{2,3}$- and Fe $L_{2,3}$-edges were measured in the TEY and PFY mode using an electrometer (KEITHLEY 6514) and a silicon drift detector (SDD, Amptek), respectively.. The $x$-controlled Na$_x$Co[Fe(CN)$_6$]$_{0.9}$ films were inserted into a vacuum chamber with a base pressure of $6 \times 10^{-8}$ Torr. CoO and Fe$_2$O$_3$ were measured as a reference for relative energy calibration. The energy resolutions were approximately 0.3 eV. The measurement was performed at room temperature.

**RIXS around the Co $L_3$-edge**

The RIXS measurements were conducted using the AGM-AGS spectrometer at TLS BL05A1 beamline[18] at the NSRRC in Taiwan. The $x$-controlled Na$_x$Co[Fe(CN)$_6$]$_{0.9}$ films were inserted into a vacuum chamber with a base pressure of $2 \times 10^{-8}$ Torr. The incident photon energy was set around the $L_3$-edge of Co. The scattering angle defined as the angle between the incident and the scattered X-rays was 90°, and the incident angle from the surface plane was 20°. The incoming X-ray was linearly polarized with the polarization perpendicular to the scattering plane, i.e., σ polarization. The beam size of incident X-ray projected on the sample was about 0.4 mm (vertical) ×0.8 mm (horizontal). Since the injection angle is 20° off the sample surface, horizontal projection of the beam on the sample was extended to about 2 mm. The RIXS spectra were recorded with a CCD detector. To avoid the radiation damage, the incident X-ray flux was reduced by narrowing vertical slit of the monochromator and the sample position was vertically shifted by 0.2 mm at every 15 min. We confirmed that the spectral profile does not change in this radiation condition. Each spectrum was recorded for about 2 hour. The total RIXS energy resolution was ~100 meV at 780 eV. The measurement was performed at room temperature.

**Acknowledgments**

This work was supported by JSPS KAKENHI (Grant Number JP17H0113 and JP16K20940). Preliminary XAS experiments were performed at the beamline BL12 of the SAGA Light Source with the approval of the Kyushu Synchrotron Light Research Center (Proposal No. 1604015R). The XAS measurements around Co $L_{2,3}$- and Fe $L_{2,3}$-edges were performed at BL08B1 beamlines at the NSRRC in Taiwan. The RIXS measurements around Co $L_3$-edge were performed at BL05A1 beamline at the NSRRC in Taiwan. The elementary analyses were performed at the Chemical Analysis Division, Research Facility Center for Science and Engineering, University of Tsukuba, Tsukuba, Japan.

**Author Contributions**

H. N performed XAS measurements, RIXS measurements, and careful RIXS analyses. M. T. prepared and characterized the $x$-controlled Na$_x$Co[Fe(CN)$_6$]$_{0.9}$ films. J. O and W. W. collaborated the XAS and RIXS measurements as staff members at the NSRRC. Y.-Y. C. and A. S. supported the launch of the XAS experiment. D.-J. H. led the XAS and RIXS team at the NSRRC. Y. M planned the research and wrote the manuscript.

**Additional Information**

There are no competing financial interests.